# A spin ladder compound doubles its superconducting $T_C$ under a gentle uniaxial pressure

D. MOHAN RADHEEP[1], R. THIYAGARJAN[1], S. ESAKKIMUTHU[1], GUOCHU DENG[2,4], E. POMJAKUSHINA[2], C.L. PRAJAPAT[3], G. RAVIKUMAR[3], K. CONDER[2,] G. BASKARAN[5,6] AND S. ARUMUGAM[1,*]

[1] Centre for High Pressure Research, School of Physics, Bharathidasan University, Tiruchirappalli 620024, India

[2] Laboratory for Developments and Methods, Paul Scherrer Institute, CH 5232 Villigen, Switzerland

[3] Technical Physics Division, Bhabha Atomic Research Centre, Mumbai 400085, India.

[4] Bragg Institute, Australian Nuclear Science and Technology Organization, 2234 Menai, Australia

[5] The Institute of Mathematical Science, Taramani, Chennai 600113, India

[6] Perimeter Institute of Theoretical Physics, Waterloo, Ontario, Canada.

* e-mail: sarumugam1963@yahoo.com

## Abstract

Discovery of new high $T_C$ superconductors, with $T_C > 23$ K, continues to be challenging. We have doubled the existing $T_C$ of single crystal $Sr_3Ca_{11}Cu_{24}O_{41}$, a spin ladder cuprate, from *12K* to *24K*, using a gentle uniaxial pressure ~ *0.06 GPa*. In contrast, earlier works used a nearly 100 times larger hydrostatic pressure ~5

**$GPa$, only to reach a maximum $T_C \sim 12K$. Our work exposes large and nearly equal, but opposing contributions to changes in $T_C$, arising from compressions along and perpendicular to ladder planes, in hydrostatic pressure experiments. In our resistivity measurements, uniaxial pressure applied along ladder planes increase $T_C$, while that perpendicular to ladder planes decrease $T_C$. Our findings *i)* offers a new hope for further increase in $T_C$ in spin ladder compounds and *ii)* calls for a large shift in phase boundaries of the currently accepted pressure-temperature phase diagram.**

A landmark discovery by Bednorz and Muller in 1986 marks the beginning of the field of high $T_C$ superconductivity. A record $T_C \sim 23\ K$, held by $Nb_3Ge$ for decades, was overtaken by doped $La_2CuO_4$, a layered cuprate. Now, superconductors with $T_C > 23\ K$ are generally called high $T_C$ superconductors (*HTSC*) [1]. Continuing attempts to find new *HTSC* compounds have yielded fruits in the form of $MgB_2$, $Ba_{1-x}K_xBiO_3$, Fullerites and recently *Fe* pnictides. Layered quasi two-dimensional cuprate family also has grown in size and stature. This family holds a maximum reproducible superconducting $T_C$ of *163 K* in certain trilayer cuprates.

A new family, a layered but quasi one dimensional cuprate, called spin-ladder compound $Sr_{14-x}Ca_xCu_{24}O_{41}$ appeared in the scene about seventeen years ago. The existence of superconductivity in the doped spin ladder system was theoretically predicted by *Dagotto et al* [2-5]. There was a dual purpose in the study of this system, *i)* to make the mechanism of superconductivity and role played by valence bond singlets in *HTSC* cuprates more transparent and *ii)* to open a route to new family of high $T_C$ superconductors. The first purpose has been partly achieved. As for the second purpose, so far spin ladder compounds have resisted crossing the high Tc boundary of 23 K.

In the present paper we report our success in elevating a spin ladder compound $Sr_3Ca_{11}Cu_{24}O_{41}$ *(SCCO)* to the status of a high $T_C$ superconductor, under a gentle uniaxial pressure of *0.06 GPa*. We have doubled the $T_C$ to *24 K*, from the known value of *12 K*. The uniaxial pressure needed to achieve this is nearly 2 orders of magnitude smaller than the hydrostatic pressure of *5 GPa*, which could only manage a $T_C \sim 12\ K$ in the past. We locate superconducting $T_C$ from the onset of resistivity drop in our electrical transport measurement. Our work is significant in the sense it offers a new hope for further increase in superconducting $T_C$ and also calls for detailed theoretical and experimental study of the superconducting state. Further, our work is a record breaking one from the point of view of effect of uniaxial pressure, exceeding earlier value of ($\delta T_C / T_C$) in $La_{1.48}Nd_{0.4}Sr_{0.12}CuO_4$ [6] and epitaxial strain induced increase in $T_C$ in $La_{1.9}Sr_{0.1}CuO_4$ [7]. Our work brings out a singular merit of uniaxial pressure over hydrostatic pressure. Hydrostatic pressure can hide, through cancellations, potentially large changes in $T_C$.

Experimentally superconductivity was first found by *Uehara et al* [8], who discovered superconductivity at high pressure in the doped polycrystalline compound $Sr_{0.4}Ca_{13.6}Cu_{24}O_{41}$ with maximum $T_C$ at about *12 K* in a very narrow pressure range *(3.0 - 4.5 GPa)*. Subsequent works, including a recent work by *Vanishri et al* [9]., have addressed several issues, including existence of an antiferromagnetic order in the *P-T* phase diagram. It is surprising that till now, studies of transport properties under direct uniaxial pressure have been neglected in this system. Thus, one of the aims of the present study was to test the sensitiveness of the spin ladder structure at different crystallographic directions to the strain/uniaxial pressure and the subsequent effect on the transport and magnetic properties for the high Ca-doped $Sr_3Ca_{11}Cu_{24}O_{41}$ system. .

Intense investigations led to a realization of various spin ladder materials such as $Sr_{14-x}Ca_xCu_{24}O_{41}$ [10, 11], $SrCu_2O_3$ [12], $Sr_2Cu_3O_5$ [13], $LaCuO_{2.5}$ [14], $CaCu_2O_3$ [15], $Bi(Cu_{1-x}Zn_x)_2PO_6$ [16] and $NaV_2O_5$ [17]. In this group of ladder compounds, $Sr_{1-x}Ca_xCu_{24}O_{41}$ is the only one known to exhibit superconductivity at low temperatures and under high pressures.

The spin ladders cuprate $Sr_{14-x}Ca_xCu_{24}O_{41}$ can be regarded as low-dimensional relative of the high-$T_C$ materials. A ladder is formed by two copper-oxide chains that are covalently coupled by oxygen atoms. Ladders are further coupled covalently, in a staggered fashion to form a plane, resulting in an anisotropic two-dimensional electronic system. The ladder compounds build a bridge between quasi 1D and 2D systems. $Sr_{14}Cu_{24}O_{41}$ (SCO) has, in addition to ladder planes, $Cu$-oxygen chains lying between ladder planes. A schematic view of the structure of the *SCCO* spin ladder compound and $Cu_2O_3$ ladder and $CuO_2$ chain layers are shown in Figure *1 (a) (b)* and *(c)*, respectively. Hybridization of the $Cu^{3d}$ and $O^{2p}$ orbitals are schematically shown in Fig. *1(b)* and *1(c)*. The crystal structure of *SCCO* (Fig. *1(a)*) shows displacements of the atoms from ideal positions (especially observed for the chain oxygen) as the structure is incommensurately modulated along *b*-axis [10, 11, 18].

In the stoichiometric compound $Sr_{14}Cu_{24}O_{41}$, there are 6 holes per formula unit, residing nominally in the $Cu^{2+}$ subsystem. It is believed that about 5 holes reside in the chain and one resides in the ladder plane. These holes are likely to be spatially ordered at low temperatures due to coulomb repulsion among themselves.

Replacement of *Sr* by isovalent *Ca* with a smaller ionic radius results in lattice distortion and chemical pressure effects. Thermal and quantum melting of holes and hole transfer between chains and ladder planes control electrical and magnetic properties of an otherwise Mott insulating ladder planes as a function of *x* (Ca doping). Hopping matrix elements $t_l$ within a ladder are large and comparable to that in the $CuO_2$ planes of cuprates, because of similar quantum chemistry. Consequently superexchanges $J_l$, within a ladder are strong, antiferromagnetic and similar in strength to that in cuprates. Inter chain and inter ladder electron hopping matrix elements $t_{cc}$ and $t_{ll}$ are smaller; and $t_l \gg t_{ll} \gg t_{cc}$. Depending on *x*, pressure and temperature holes can be *i)* frozen in the ladders, *ii)* delocalized within the ladder and *iii)* delocalized within the ladder plane. Thus we have possibility of a rich physics of localization, *1D* and *2D* transport.

A strong singlet pairing within the ladders (rung singlets) can cause quasi two dimensional superconductivity, through Josephson coupling or pair tunneling between ladders within a given ladder plane. In addition to standard superconductivity, a pair density wave, a finite momentum condensation of Cooper pairs (often referred to as hole crystal) is also possible. There are also suggestions that chains and planes could support long range antiferromagnetic order, once holes crystallize [19].

Hydrostatic and uniaxial pressures are important tools in solid state physics. Unlike hydrostatic pressure, uniaxial strain is capable of distinguishing changes in physical properties, when applied selectively along principal directions such as *a, b* axis of the $CuO_2$ *ab*-plane or *c*-axis of cuprates. In the case of cuprates it is known that large changes of superconducting $T_C$ occur in opposite directions, when *ab*-plane or *c*-axis are compressed separately [20]. Thus hydrostatic pressure often hides these large changes, arising from cancellation and opposing trends. Recent theoretical work attempts to quantify changes in superconducting $T_C$ under Uniaxial pressure in planar cuprates [21].

Inspite of a large body of experimental and theoretical work, some important questions remain unsettled. A major question is the hole content in the ladder plans at low temperatures as a function of *x*, *Ca* substitution. Interpretation of different experimental results seem to give conflicting information about the hole density in the ladder plane. However, it is widely believed that hole content is low and does not show marked variation as a function of *x* [19].

The present work was partly inspired by the effect of the uniaxial pressure (*P ab*-plane) found in $La_{1.48}Nd_{0.4}Sr_{0.12}CuO_4$ system where simultaneous suppression of the insulating low temperature tetragonal stripe phase and an appearance of superconductivity with a drastic increase in $T_C$ was already observed at a very low pressure of *0.04GPa* [6]. Hence, we made systematic investigations of the temperature dependence of the resistivity on the crystallographically oriented single crystals of $Sr_3Ca_{11}Cu_{24}O_{41}$ under direct uniaxial pressure *(< 0.05GPa)* with *P // a ($P_a$), P // b ($P_b$)* and *P // c ($P_c$)*. Consequently, superconductivity was induced applying small (< *0.05GPa*) uniaxial pressure in $P_c$ and $P_a$ directions.

**Experimental Methods**

In this study single crystal samples of $Sr_3Ca_{11}Cu_{24}O_{41}$ were grown using Traveling Solvent Floating Zone (*TSFZ*) method [22]. The quality of the single crystals grown was checked by Laue camera, X-ray diffraction (D8-Brucker), neutron diffraction *(Morpheus)* and polarized microscope at Paul Scherrer Institute. The crystals consisted of single homogeneous domain and were phase pure. The crystals were oriented along *a*, *b* and *c* directions using Laue X-ray method and cube-shaped pieces with edge lengths of *1.5-2.5 mm* were cut with a wire-saw without any change in the alignment. The crystals were characterized by resistivity measurements using standard four probe method with *4K* Closed cycle Refrigerator (CCR) and variable temperature insert (VTI) from *4K-300K* (Cryo Mech, USA). The electrical resistivity measurements under uniaxial pressure were carried out similarly on those previously reported by *Arumugam et al* [23] with a little modification to suit into CCR-VTI. The field cooled magnetization measurements were carried out at ambient pressure and under uniaxial pressure using clamp type *Be-Cu* uniaxial pressure cell and Magnetic Property Measurements System MPMS (Quantum Design, USA) system with the magnetic field of *100 Oe*. The clamped pressure was calculated from the known applied force and the area of the sample using torque wrench. It was also rechecked by measuring the superconducting transition temperature of *Sn* with various applied forces by keeping the area of the sample and spring constant. Further details are reported elsewhere [24].

## Results and Discussion

Pressure was applied parallel to *c*-axis ($P_c$), *a*-axis ($P_a$) and *b*-axis ($P_b$) and resistivity was measured along *a*-axis ($\rho_a$), *c*-axis ($\rho_c$) and *a*-axis ($\rho_a$)-axis, respectively, for various applied pressures upto 0.06 *GPa* in the temperature range *4-150K*. Results are shown in Fig. *2(a), 2(b)* and *2(c),* respectively. The inset of Fig. *2(a)* shows the temperature dependence of $\rho_a$ for $P_c$ configuration up to *30 K*. The directions of the applied pressure and the measured resistivity are shown on corresponding insets in Fig. 2. It was found that the resistivity ($\rho_a$, $\rho_c$, and $\rho_a$) at ambient pressure did not appreciably depend on temperature down to *50 K*. At lower temperatures, however, resistivity increased rapidly indicating a semiconducting nature of the material. In case of $\rho_a$, the low temperature upturn disappeared at *0.042 GPa* and a further increase of the pressure led to the appearance of superconductivity with $T_C$ of ~5K at *0.045 GPa*. As shown in Fig.*3*, it was observed that $T_C$ increases as the pressure is growing with a corresponding pressure coefficient $dT_C/dP_c$~ *1130 K/GPa*. It is to be noted that the $T_C$-values were acquired from the derivative plots. Similarly, pressure was applied in $P_a$ direction and the temperature dependence of $\rho_c$ was measured for various pressures as shown in Fig. *2(b)*. An indication for superconductivity was observed only at a pressure of *0.049 GPa* with $T_C$~6.5K.

Finally the pressure was applied in $P_b$ direction and the temperature dependence (*4-150 K*) of $\rho_a$ was measured for various pressures upto *0.048 GPa*. As evidenced on Fig.*2(c)*, a rapid increase of resistivity was observed close to *4 K*. The resistivity upturn trend at low temperatures increased with a pressure. Applying pressure along *b* direction favored an insulating state while both others configurations $P_a$ and $P_c$ favored a superconducting state. Many attempts which were carried out in order to measure resistivity at higher pressures failed as the crystals broke easily under uniaxial pressure independently on the applied pressure directions [23]. It has to be stressed that the uniaxial pressure caused much larger and anisotropic mechanical stress in the samples in comparison with the hydrostatic one.

Magnetization measurements for $Sr_3Ca_{11}Cu_{24}O_{41}$ crystals were performed using MPMS in order to verify resistivity measurements and to estimate superconducting volume fraction. Measurements were performed in a clamp type uniaxial pressure cell. Temperature dependence of the magnetization was measured in the field cooled regime at various pressures in $P_a$ direction with the magnetic field of *100 G*. As it is shown in Fig. *4* superconducting transition was observed at ~ *6.8 K* already applying a pressure of *0.05 GPa*. The volume fraction of superconductivity at *5 K* was estimated to be ~ *7%*. The obtained value matches with the superconducting volume fraction of ~ *5%* at *4.2K* and *3.5 GPa* obtained under hydrostatic pressure in polycrystalline $Ca_{13.6}Sr_{0.4}Cu_{24}O_{41}$ [25-27]. We found that the superconducting volume fraction increased slightly with a little increase in applied pressure. We were unable to do measurements more than *0.05 GPa* under uniaxial pressure due to the structural nature and breaking of the crystals frequently.

The obtained results clearly show that the spin ladder *SCCO* structure is more sensitive in the $P_c$ direction along legs than in the $P_a$ -along the rungs. The strain induced under relatively small uniaxial pressure ($P_a$, $P_c$) induces increase in the charge carrier (hole) concentration and/or delocalizes holes within the ladder plane.

It has been reported that the compressive epitaxial strain induced a large increase of $T_C$ (*49 K*) in thin films of $La_{1.9}Sr_{0.1}CuO_4$ superconductors in which the normal state resistivity changes from insulator to metallic [7]. This is comparable with our uniaxial pressure results in the spin ladder systems. The *Ca* doping induced a chemical pressure and the additional uniaxial strain in $P_a$ and $P_c$ directions similar to an external stress, which drastically affected *Cu-O* bond length and *O-Cu-O* bond angle. Recently, *Guochu et al* [18] reported that the increase of *Ca* doping has the effect of *Cu-O* bond lengths and *O-Cu-O* angles on the chain and ladders and that it increases the interaction between the two sublattices. *Guochu et al* [18] and *Isobe et al* [28] reported an evidence for a charge transfer from the chains to the ladders with an increase of doping from bond valance sum calculations.

We strongly believe that it would be possible to get a clearer understanding of the mechanism of pressure induced superconductivity in the future performing structural analysis, neutron diffraction, optical, muon spin rotation and NMR measurements of various parents and doped spin ladder systems under uniaxial pressure at low temperature. It is important and also of interest to search for superconductivity under uniaxial pressure in the $Sr_{14-x}Ca_xCu_{24}O_{41}$ system in the whole $0 \leq x \leq 14$ doping range and also other parent and doped spin ladder systems such as $CaCu_2O_3$, $SrCu_2O_3$, $Sr_3Cu_2O_5$, $BiCu_2PO_6$, $NaV_2O_5$, and $Ni_3V_2O_8$ etc.

Within our experimental constraints it is difficult to go below *0.04 GPa* to locate onset of superconductivity as a function of uniaxial pressure and determine nature of the quantum phase transition. As we have an insulator to metal transition, long range coulomb interactions is likely to make a first order phase transition. This is corroborated by the work of *Vanishri et al.,* [9] where they observe a phase coexistence region of superconductivity and antiferromagnetism, a consequence of first order transition, as a function of hydrostatic pressure, albeit at a much higher pressure, It is likely that in our system $Sr_3Ca_{11}Cu_{24}O_{41}$, insulating state to the left to the superconductor phase is an antiferromagnetic or a quantum spin liquid state or a pair density wave (Inset figure (5)).

Based on available experimental work and theoretical inputs a couple of phase diagram has been suggested in the variable *x, P* and *T* in the literature. In figure (5), we present our phase diagram focusing on superconductivity. A quantitative and large shift in the phase boundary is manifest in the log scale. From our work it is clear that hydrostatic pressure does not expose a large increase in $T_C$ that is caused by compression along the ladder axis. It is compensated or cancelled by an opposing effect in $T_C$ arising from compression perpendicular the ladder plane. Figure (5) gives a schematic modified phase diagram, in the light of our present finding.

In trilayer *Hg* and *Tl* cuprates $T_C$ increases up to the record high value of *163 K* under pressure. Does this mean that we could achieve a $T_C$ higher than the *24 K* we have observed in the present experiment? Superexchange *J*, a key factor that provides glue for Cooper pairing or singlet formation, continues to be large within the ladders, as in $CuO_2$ planes. *J* is a robust quantum chemical parameter of the strongly hybridized Cu and oxygen square planar complex, as evidenced by theoretical calculation and a variety of experiments. *Thus local superconducting $T_C$ scale or local pairing amplitude within a given ladder continues to be as large as that in cuprates with a high $T_C$. A sufficiently strong Josephson coupling or inter ladder pair tunneling should stabilize a larger $T_C$ than that has been observed so for.*

Weakly coupled ladders, unlike planar cuprates, has no special affinity for stripe orders, which is a major degrader of superconducting $T_C$ in planar cuprates. Orders that competes with superconductivity are hole localization or hole ordering within ladders. In other words, an optimally doped and weakly coupled ladder with no competing hole localization should be capable of creating a higher $T_C$ than what we have observed. At the moment there is no strong suggestion from either experiment or theory that can precisely determine hole density in the ladder plane. Reaching an optimal hole density in the ladder plane to reach a higher $T_C$ is a future possibility that our work suggests. It will be also interesting to mimic uniaxial pressure through epitaxial strain and confirm our doubling of $T_C$ and go beyond.

Now that spin ladder compound has just entered high $T_C$ regime, it is of interest to find symmetry of the superconducting order parameter. From theory point of view, within the ladder we have a d-wave order parameter. Symmetry of the superconducting order parameter stabilized by the weak inter-ladder coupling needs to be studied in detail both in theory and experiments.

We conclude with three optimistic notes: *i)* there is room for further increase in $T_C$ in the spin ladder compounds using Uniaxial pressure *ii)* the large but cancelling changes in $T_C$ caused in hydrostatic pressure that we have identified gives us a hope that uniaxial pressure experiments in the trilayer cuprates might help us exceed the current record $T_C$ of *163 K*, obtained by hydrostatic pressure, and go closer to room temperature scales and *iii)* it is possible to observe Uniaxial pressure induced superconductivity in other existing spin ladder systems.

**Acknowledgement**

The author SA wishes to thank DST (SERC, Indo-Swiss and IDP) and UGC for the financial support to carry out the research work. The author KC acknowledges for Indo-Swiss Joint Research Program (ISJRP, Contract No. JRP122960) by Swiss State Secretariat of Education and Research. The author DMR would like to thank CSIR-SRF fellowship scheme for providing financial support. We acknowledge useful discussions with Dr. Michel Kenzelmann, PSI and Dr. Henrik Rønnow, EPFL, Switzerland. This research was supported by Perimeter Institute for Theoretical Physics.

**Author Contributions**

The author SA are envisaging the problem, characterization of single crystals through transport and magnetic measurements, the development of uniaxial pressure device and measurement of electrical resistivity and magnetization under uniaxial pressure. The author DMR is involved in crystal growth, chemical and structural characterization, orienting and cutting of the crystals, measurement of electrical resistivity and magnetization under uniaxial pressure. The authors GD, EP and KC are involved in syntheses, crystal growth, chemical and structural characterization, orienting and cutting of the crystals. G Baskaran contributed to placing the result in the theoretical frame work of spin ladder physics and also help bring out the first order nature of the phase transition. The author RT is involved in crystal growth, chemical and structural characterization and Orienting and cutting of the crystals. The author SE, CLP and GR are involved in measurement of magnetization under uniaxial pressure. All authors contributed to the writing of the paper.

**Figure Legends:**

**Figure 1:** (a) The structure of the cuprate family $Sr_{14-x}Ca_xCu_{24}O_{41}$. Layers of the constituents $(Sr^{2+}, Ca^{2+})$ are alternate with the chain and the ladder layers; (b, c) Schematic diagram of the hybridization of the *Cu3d* and *O2P* orbitals within the ladder and chain layers and antiferromagnetic *(AF)* and ferromagnetic *(FM)* coupling are shown.

**Figure 2:** Temperature dependence of electrical resistivities of $Sr_3Ca_{11}Cu_{24}O_{41}$ single crystal under various uniaxial pressures (a) across the ladder in the ladder plane $(\rho_a)$ with $P \parallel c$-axis (b) along the ladder plane $(\rho_c)$ with $P \parallel a$-axis (c) across the ladder in the ladder plane $(\rho_a)$ with $P \parallel b$-axis [Inset 2a : $\rho_a$ vs $T$ with $P \parallel c$-axis upto *30 K*]

**Figure 3:** Pressure dependence of $T_C$ determined from the electrical resistivity $(\rho_a)$ measurements with $P \parallel c$-axis of $Sr_3Ca_{11}Cu_{24}O_{41}$ single crystal. The solid line is guide to the eye.

**Figure 4:** Magnetic susceptibility $(\chi_a)$ as a function of temperature with $P \parallel a$-axis of $Sr_3Ca_{11}Cu_{24}O_{41}$ single crystal. [Inset: $\chi_b$ vs $T$ with $P \parallel b$-axis]

**Figure 5:** Modified phase diagram of $Sr_{14-x}Ca_xCu_{24}O_{41}$. [Inset: Schematic phase diagram of $Sr_3Ca_{11}Cu_{24}O_{41}$ under uniaxial pressure]

**Figures**

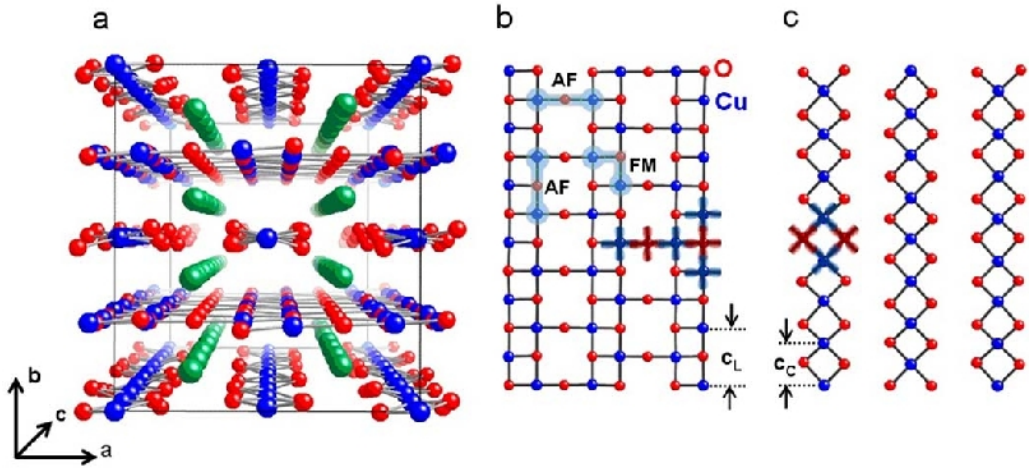

**Figure 1:** (a) The structure of the cuprate family $Sr_{14-x}Ca_xCu_{24}O_{41}$. Layers of the constituents $(Sr^{2+}, Ca^{2+})$ are alternate with the chain and the ladder layers; (b, c) Schematic diagram of the hybridization of the *Cu3d* and *O2P* orbitals within the ladder and chain layers and antiferromagnetic *(AF)* and ferromagnetic *(FM)* coupling are shown.

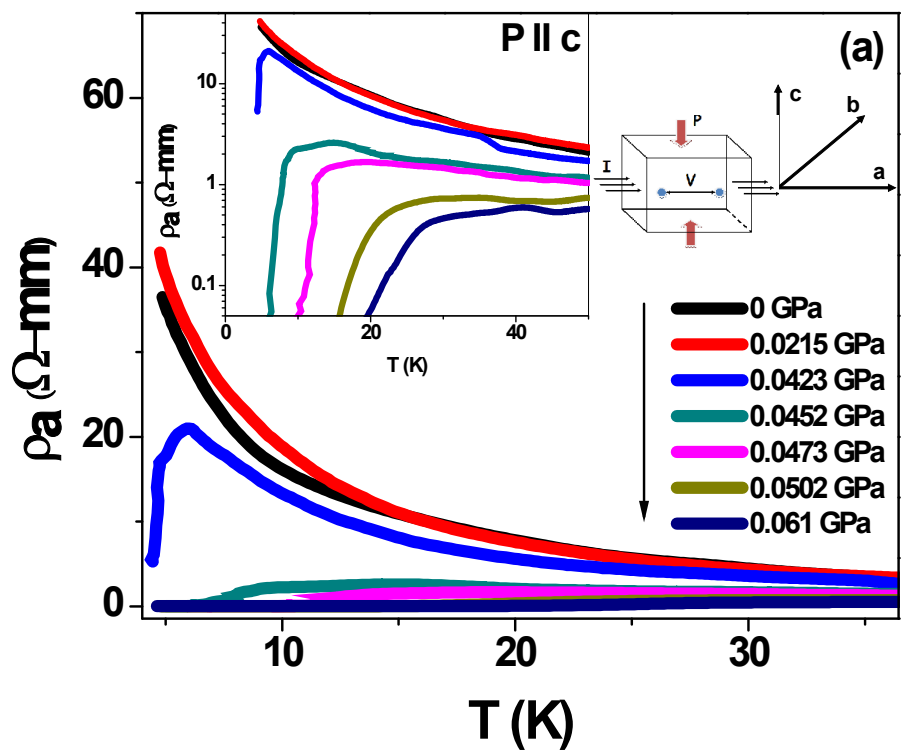

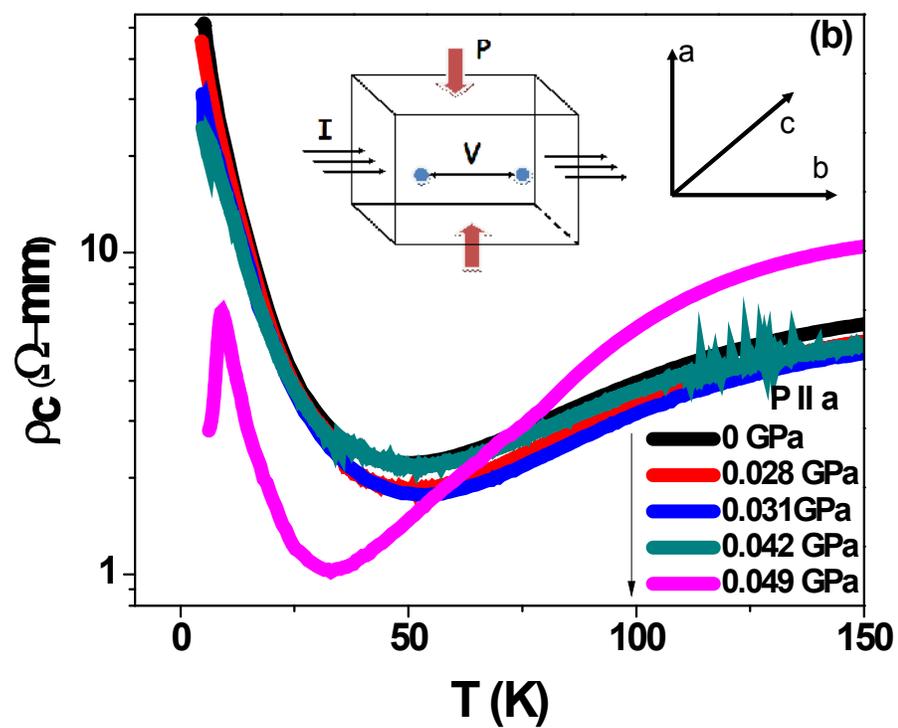

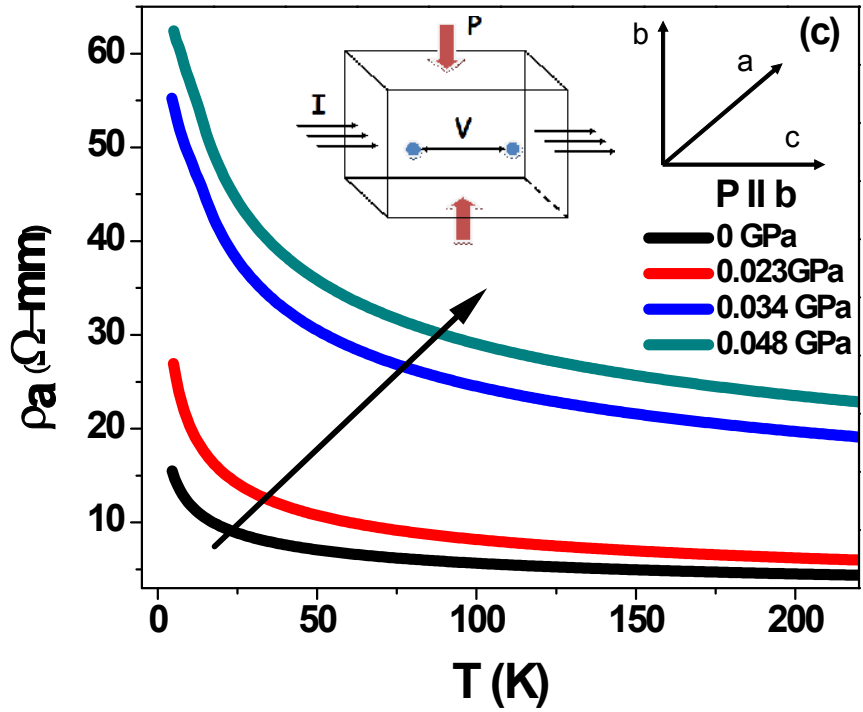

**Figure 2:** Temperature dependence of electrical resistivities of $Sr_3Ca_{11}Cu_{24}O_{41}$ single crystal under various uniaxial pressures **(a)** across the ladder in the ladder plane $(\rho_a)$ with $P \parallel c$-axis **(b)** along the ladder plane $(\rho_c)$ with $P \parallel a$-axis **(c)** across the ladder in the ladder plane $(\rho_a)$ with $P \parallel b$-axis [Inset 2a : $\rho_a$ vs T with $P \parallel c$-axis upto 30 K]

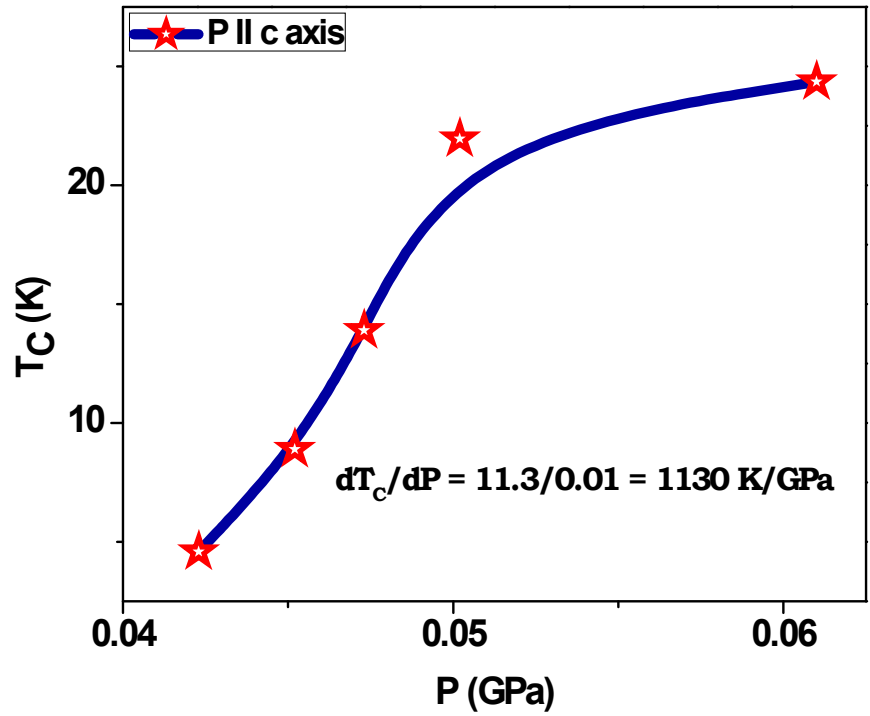

**Figure 3:** Pressure dependence of $T_C$ determined from the electrical resistivity $(\rho_a)$ measurements with $P \parallel c$-axis of $Sr_3Ca_{11}Cu_{24}O_{41}$ single crystal. The solid line is guide to the eye.

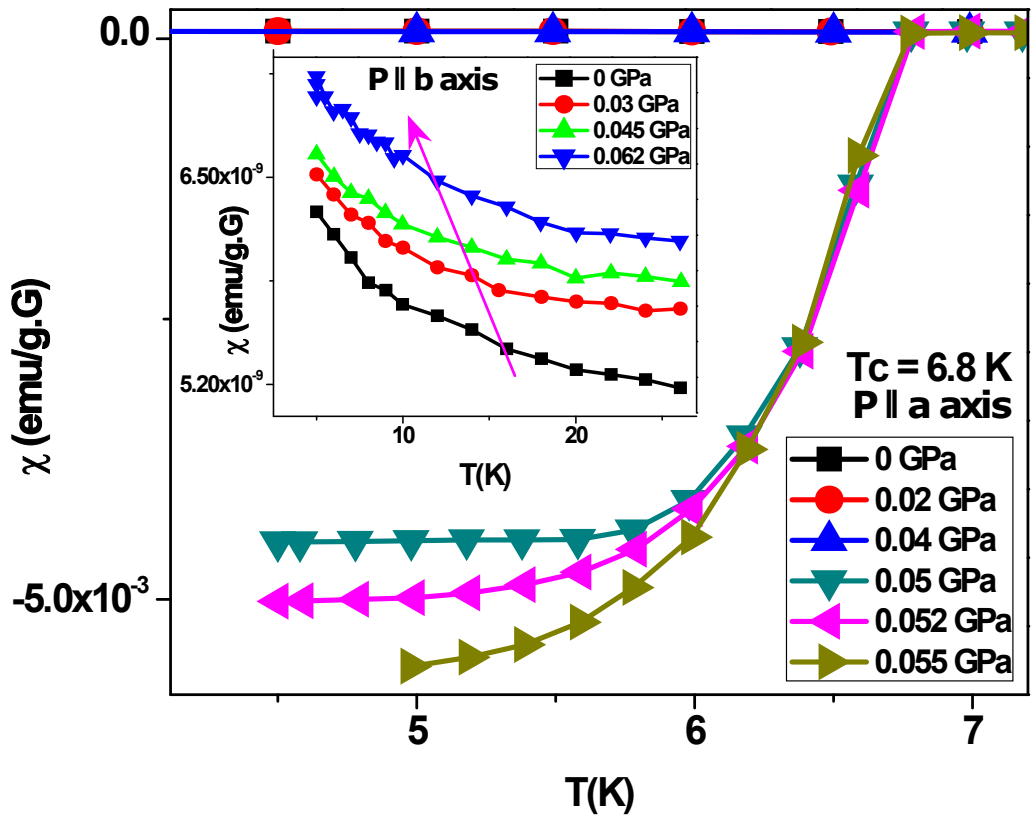

**Figure 4:** Magnetic susceptibility ($\chi_a$) as a function of temperature with $P \parallel a$-axis of $Sr_3Ca_{11}Cu_{24}O_{41}$ single crystal. [Inset: $\chi_b$ vs $T$ with $P \parallel b$-axis]

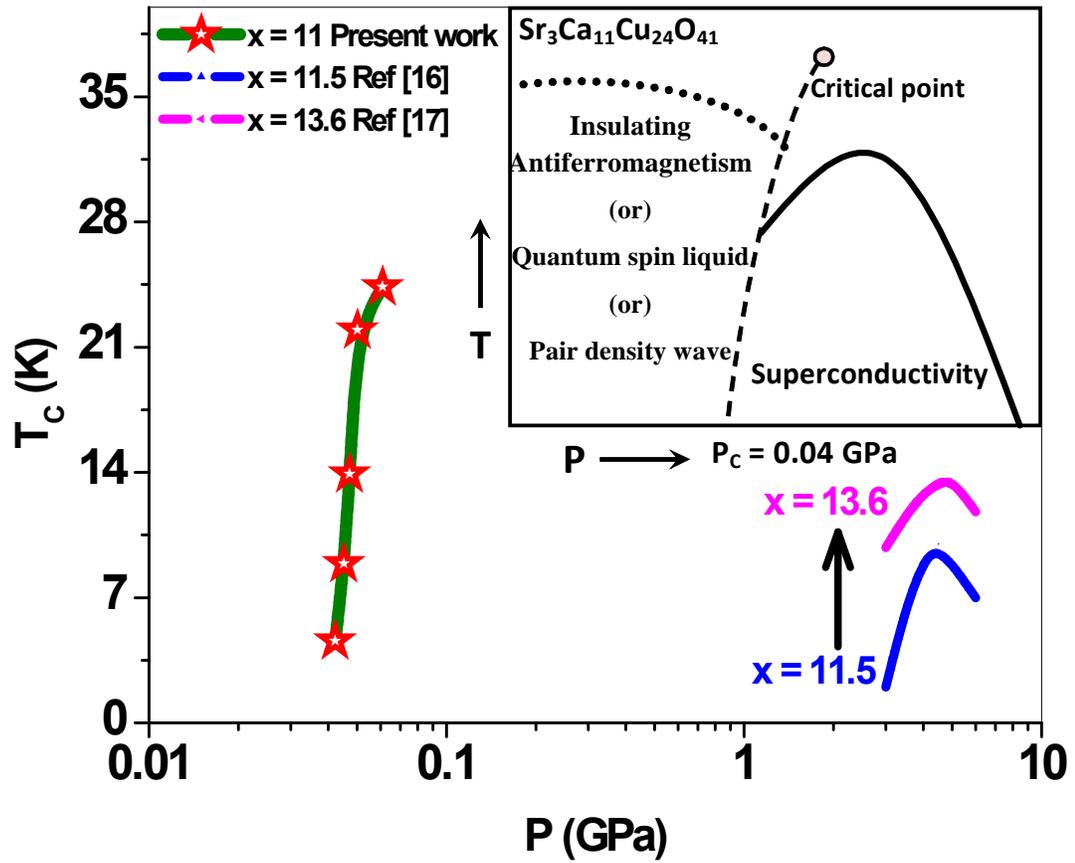

**Figure 5:** Modified phase diagram of $Sr_{14-x}Ca_xCu_{24}O_{41}$. Inset : Schematic phase diagram of $Sr_3Ca_{11}Cu_{24}O_{41}$ under uniaxial pressure